\documentclass[12pt,preprint]{aastex6}

\usepackage{float}

\shorttitle{Montage and Visualization}
\shortauthors{Berriman and Good}

\begin{document}


\title{The Application of the Montage Image Mosaic Engine To The Visualization Of Astronomical Images}


\author{G. Bruce Berriman\altaffilmark{1} and J.C. Good\altaffilmark{2} }

\affil{IPAC, Mail Code 100-22, Caltech, 1200 E. California Blvd., Pasadena, CA 91125.}

\altaffiltext{1}{gbb@ipac.caltech.edu}
\altaffiltext{2}{jcg@ipac.caltech.edu}


\begin{abstract}

The Montage Image Mosaic Engine was designed as a scalable toolkit, written in C for performance and portability across *nix platforms, that assembles FITS images into mosaics. The code is freely available and has been widely used in the astronomy and IT communities for research, product generation and for developing next-generation cyber-infrastructure. Recently, it has begun to finding applicability in the field of visualization. This has come about because the toolkit design allows easy integration into scalable systems that process data for subsequent visualization in a browser or client. And it includes a visualization tool suitable for automation and for integration into Python: {\tt mViewer} creates, with a single command, complex multi-color images overlaid with coordinate displays, labels, and observation footprints, and includes an adaptive image histogram equalization method that preserves the structure of a stretched image over its dynamic range. The Montage toolkit contains functionality originally developed to support the creation and management of mosaics but which also offers value to visualization: a background rectification algorithm that reveals the faint structure in an image; and tools for creating cutout and downsampled versions of large images. Version 5 of Montage offers support for visualizing data written in HEALPix sky-tessellation scheme, and functionality for processing and organizing images to comply with the TOAST sky-tessellation scheme required for consumption by the World Wide Telescope (WWT). Four online tutorials enable readers to reproduce and extend all the visualizations presented in this paper.

\end{abstract}


\section{Introduction}

Desktop tools such as SAOImage-DS9 (\cite{joma03}; \url{http://ds9.si.edu/site/Home.html}), FITS liberator  (\cite{niel08}; \url{https://www.spacetelescope.org/projects/fits_liberator/}) and ESASky (\cite{meri16}; \url{http://sky.esa.int/})  are invaluable in visualizing astronomical images. They are not, however, intended for automated creation of images from large collections of data, nor for integration into pipelines or workflows \textemdash two of six visualization Grand Challenges identified by \cite{hafl11}. Because its design can support these two Grand Challenges, the Montage Image Mosaic Engine (\cite{jaco10}; \cite{berr16};  \url{http://montage.ipac.caltech.edu}; Astronomy Source Code Library record ascl:1010.036; dx.doi.org/10.5281/zenodo.49418) is finding growing applicability in the field of visualization. Originally delivered to create mosaics of images written in Flexible Image Transport System (FITS) format \citep{cagr02}, Montage is a toolkit that is easily integrated into scalable systems designed to process images for subsequent visualization. Moreover, it provides a utiliity for creating Portable Network Graphics (PNG) or Joint Photographic Experts Group (JPEG) representations of FITS images that can be used in an automated fashion. A similar automation tool, STIFF \citep{bert12}, has been used successfully by \cite{lebr15}, \cite{bail11} and \cite{mein16} and others. 

There are aspects of Montage, built as part of the process of building and managing mosaics, that also contribute value to visualization. Montage models and rectifies the sky background to a common level and thus reveals faint, diffuse features; it offers an adaptive image stretching method that preserves the full dymamic range of a FITS image; it provides utilities for creating cutouts of large images and downsampled versions of large images that can then be visualized on desktops or in browsers; and it resamples and reprojects images to a common grid and enables multi-color visualization.

Version 5.0 of Montage offers capabilites for visualizing two sky-tessellation schemes that are not readily amenable to visualization in their native forms. Data sets written as Hierarchical Equal-Area isoLatitude Pixelization (HEALPix) maps \citep{gors05} can be reprojected into all common spherical projections used in astronomy, and then visualized and and studied with other wide-area sky maps. And FITS images stored in all commonly used spherical projections can be reprojected and organized into the Tessellated Octahedral Adaptive Subdivision Transform (TOAST) sky-tessellation scheme required for consumption by the World Wide Telescope (WWT) \citep{good12}. 

This paper presents visualizations of images that exploit the above capabilities,  and contains links to four supplementary online tutorials that allow readers to recreate, adapt and extend these visualizations. 

\section{The Design and Release History of Montage}

A knowledge of the design of Montage is valuable in understanding its applicability to visualization. Montage is a toolkit for creating mosaics that preserve the calibration and astrometric fidelity of input FITS images  \citep{jaco10}. It can process two-dimensional images and data cubes \citep{berr16}. The toolkit is written in ANSI-C for performance, is portable across all common *nix platforms, accepts input from the command line, and returns structured American Standard Code for Information Interchange (ASCII) responses that can be parsed by any computer. The tools scale from desktops, where they are usually run serially through scripts, to high-performance platforms, where they are parallelized through workflow managers such as Pegasus, described in \cite{desi05} and \cite{deva16}, or through the Message Passing Interface (MPI). The code is distributed with a Berkeley Software Distribution (BSD) 3-clause license and freely available from GitHub (\url{https://github.com/Caltech-IPAC/Montage}) or the Montage website (\url{http://montage.ipac.caltech.edu/docs/download.html}). The toolkit is self-contained with all necessary support libraries, and built with a {\tt make} command. The libraries include the Smithsonian Astrophysical Observatory (SAO) WCSTools library (hereafter, WCSTools) \url{(http://tdc-www.harvard.edu/wcstools/}), which implements the World Coordinate System (WCS) transformations between pixels and spatial coordinates of images \citep{mink14}. By default, Montage is able to process all spherical image projections that are supported by WCSTools.

Montage creates mosaics in response to the user's specifications of output coordinate system, image reprojection, pixel sampling and image rotation angle. The toolkit contains components that perform the tasks needed to create such mosaics:  reprojection and resampling of the input images; rectification of the variable sky and instrumental background across the images to a common level; and co-addition of the reprojected and rectified images. It also contains utilities for performing tasks such as managing large-scale mosaics and analyzing the metadata of FITS files for content and completeness.

The functionality has evolved from its first release in 2003. Versions 1 to 3 (2003-2010) offered aggregation of two-dimensional images into mosaics, and version 4 (2015) supported the same functionality for multi-dimensional image data sets (hereafter,``data cubes" for simplicity). Version 5 (2016) supports processing of HEALPix data, now the standard format for storing wide-area cosmic-background data sets, and TOAST, required for presentation of images in the WWT. Montage takes the approach of treating these two sky-tessellation schemes as WCS projections \citep{calro07} so that all the functionality in Montage is accessible to them. Altogether, there have been over 20,000 downloads to date of the various releases. 

\section{Incorporation of Montage into Visualization Environments}
Users have taken advantage of the toolkit design to perform research on their local machines and clusters, and to integrate it into workflows and pipelines that create new data products.  Recently, it has begun to find new applicability in integration within visualization environments. These environments primarily exploit the Montage functionality to co-register images measured at different epochs and at different wavelengths, which is a common use case in scientific analysis of images; e.g. \cite{hard16}, \cite{bois16}, \cite{davi16}, and \cite{kim16}.

The growth in volume in modern astronomy datasets is driving the development of large-scale image processing on remote servers, for presentation in a client application or a browser. VisiOMatic \citep{bepm15} and Toyz \citep{moma15} are two instances of this approach. \cite{luci14} integrated Montage into a client-server architecture intended as a demonstration of how a visualization environment would operate when extended to petascale processing. In this architecture, Montage plays the role of reading the images, co-registering them, re-writing them to a grid, and then creating JPEG images for visualization in a browser. The system creates a library of indices of the images based on geohashing to enable fast location and overlays of images, which are then sent to a web browser for visualization. 

\cite{mand16} has taken a different approach in developing JS9, a web-based analog of the DS9 desktop visualizer (\url{http://js9.si.edu/}). He used the {\tt mProjectPP} module, dedicated to fast reprojection of images in tangent plane reprojections, for processing images in the browser itself. Specifically, he incorporated  {\tt mProjectPP} into a prototype image blending function, in accord with the imaging compositing and blending rules proposed by the World Wide Web Consortium (W3C). It presents images from then Chandra X-ray mission, the Spitzer Space Telescope and the Galaxy Evolution Explorer (GALEX), and when JS9 is complete, users will be able to upload and blend their own images.

\cite{vogt16} used Montage to create demonstration images showing the gas content of HI gas galaxies as part of their study of the applicability of the eXtensible 3D (X3D) file format in publishing and printing three-dimensional images. Montage has been integrated into the Astronomical Plotting Library in Python (APLpy; \url{https://aplpy.github.io/}), which produces publication-quality plots of astronomical imaging data in FITS format. Montage is used to underpin image compositing services and creating multi-color images.

\cite{fern15} have stated how pre-computing mosaics with Montage can be valuable in providing the best quality data for input to the Hierarchical Progressive Surveys (HiPS) data organization scheme for managing wide-area data sets. This scheme, a proposed International Virtual Observatory (IVOA) standard, is based on HEALPix \citep{gors05} and represents a generic method of packaging, storing and describing asttronomical data. It enables progressive visualization of data sets through tools such as Aladin and ESASky. Indeed, processing of one of the data sets described in detail by \cite{fern15}, GLIMPSE360, used Montage as a mosaic and background rectification engine \citep{mead14}. A current working draft of the HiPS standard is available at \url{http://www.ivoa.net/documents/HIPS/20160623/}.

\section{Visualization Tools in the Montage Toolkit}
Versions 3 and earlier contained a visualization utility, {\tt mJPEG}, that created JPEG representations of FITS images and provided flexibility in stretching the images for display. It was developed to enable bulk creation of images and has been used as such to create preview images for, e.g., the Palomar Transient Factory \citep{lahe14} and the Starbirds data set \citep{mcqu15}. Version 4 includes {\tt mViewer}, which extends the functionality of {\tt mJPEG} and creates PNG representations of FITS images, allows full-color (three-image) displays of images with optional color enhancement, integration with Python, and image overlays as follows:

\begin {itemize}

\item Coordinate grid overlays (any coordinate system, including Besselian/Julian and Equinox selection);
\item Astronomical catalog overlays with data in any coordinate system; 
\item Multiple symbols and any color; 
\item Optional scaling of the symbols by flux or magnitude ; 
\item Image metadata (footprints) overlays, through interpreting the WCS keywords or through reading the positions of the four corners of the image;
\item Custom markers and labels.

\end{itemize}

\begin{figure} 
    \centering
    \includegraphics[width=6in]{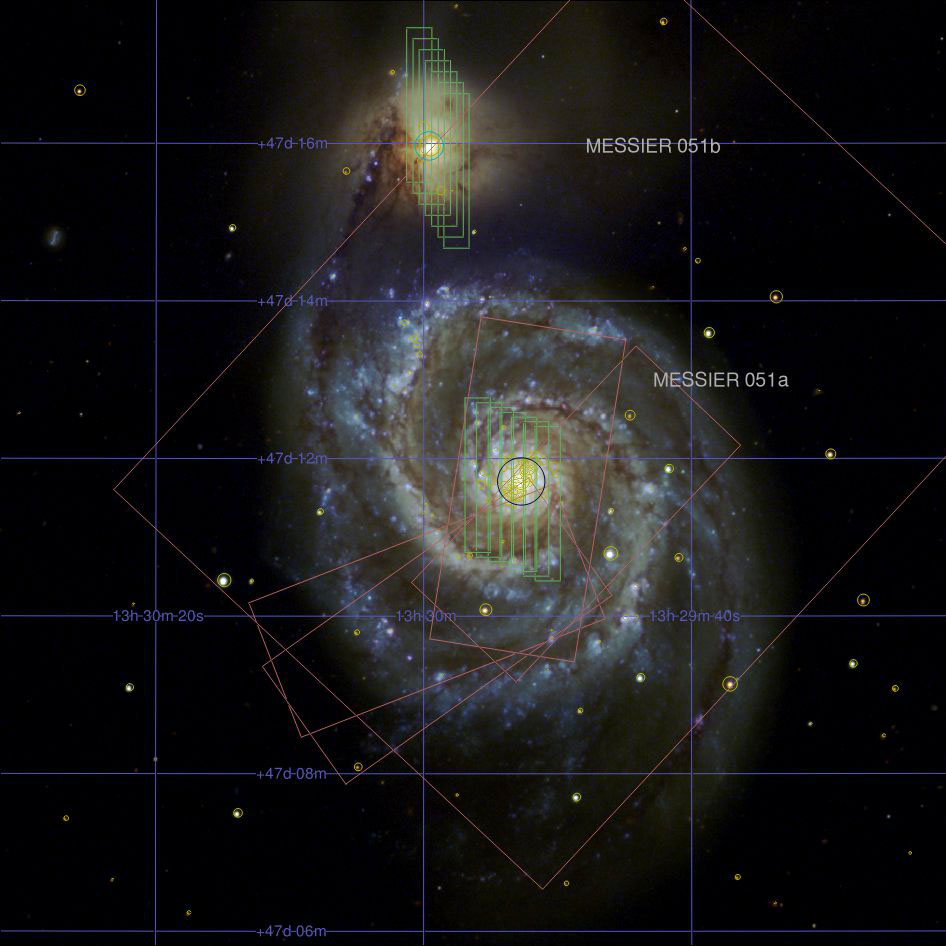}
    \caption{A three-color mosaic of M51 in the  u, g, and r bands of the Sloan Digital Sky Survey (SDSS), shown with an Equatorial J2000 coordinate grid, overlaid with the positions of 2MASS point source 
catalog sources in the J-band, scaled according to brightness (yellow circles), and with footprints from the Spitzer InfraRed Spectrograph (IRS) Peak-up images (red boxes) and the Multiband Imaging Photometer for Spitzer (MIPS) Spectral Energy Distribution (SED) images (green boxes). The image was created with a single call to {\tt mViewer}, the visualization tool included in Version 4 of Montage.}
    \label{m51}
\end{figure}

Figure~\ref{m51} illustrates these capabilities with an image created with a call to {\tt mViewer} of a mosaic of M51, built from SDSS u-, g- and r-band data. 

As modern datasets often contain large images, Montage provides the {\tt mSubimage} utility to return cutouts of sections of such images in FITS format \citep{swar09}, and the {\tt mShrink} utility to return downsampled images, also in FITS format \citep{lahe14}.

All the visualizations in this paper were created with single calls to {\tt mViewer}, and illustrate the capabilties Montage brings to visualization of modern data sets. The examples are supplemented with four online tutorials, summarized in Table ~\ref{supplements}, that allow readers to reproduce and extend sample visualizations presented here, once the required version of Montage is installed on the reader's machine. The tutorials are self-contained, with links to all required data, and their texts consequently replicate some of the material in this paper. Column 5 specifies the earliest version of Montage required for each tutorial.

\begin{deluxetable}{clllc}
  \tablecolumns{5}
  \tablewidth{0pc}
  \tablecaption{\label{supplements} Summary of Supplementary Online Tutorials}
  \tablehead{
    \colhead{Number} &
    \colhead{Title} &
    \colhead{Short URL} &
    \colhead{Section} &
    \colhead{Version} \\
  }
\startdata
 1 &Creation of a Spatial Coverage Map  & \url{http://bit.ly/2cRc3Ku} & Section~\ref{Sky Coverage}  & 	4.0 \\
 2 & Visualization and Animation of a Data Cube  &  \url{http://bit.ly/2ddVdbV}  & Section~\ref{Cubes}   &  4.0 \\
 3 & Visualization of HEALPix Maps &  \url{http://bit.ly/2dZtOwe}  & Section~\ref{HE} & 	5.0 \\
 4 & Displaying Images in the WWT  &  \url{http://bit.ly/2cRcDHZ}  &  Section~\ref{Toast} &  5.0 \\
\enddata
\label{sst}
\end{deluxetable} 

\section{Background Rectification and Visualization of the Science Content of Images}

The faint astrophysical structure in a mosaic or large-format image is most effectively seen when the spatially variable sky and instrumental radiation has been removed. Montage uses a global relaxation technique that rectifies background differences between images under the assumption that the input images are all calibrated to an absolute energy scale (that is, brightnesses are absolute and should {\it not} be modified by the rectification), and that any discrepancies between the images are due to variations in their terrestrial or instrumental background levels. \cite{mein16}, \cite{pete16} and \cite{farn16}, among others, called out the value of this background rectification to their analyses. In particular, \cite{mein16} created multi-wavelength mosaics of sources in Orion A as part of the VISTA Orion A survey.  The background-rectification algorithm assumes that terrestrial and instrumental backgrounds can be described by simple functions or surfaces, such as slopes and offsets. It assumes that the ``non-sky" background has very little energy in any but the lowest spatial frequencies. Describing the backgrounds by higher-order surfaces would very likely correct the astrophysical structure present in the image, as well as the sky background. When the ``sky" includes background containing patchy ``airglow" features, such as in the Two Micron All Sky Survey (2MASS) H-band images \citep{skru06}, the algorithm cannot distinguish these from variations in the real extra-terrestrial sky, and so they are only partly rectified.

\begin{figure} 
    \centering
    \includegraphics[width=6in]{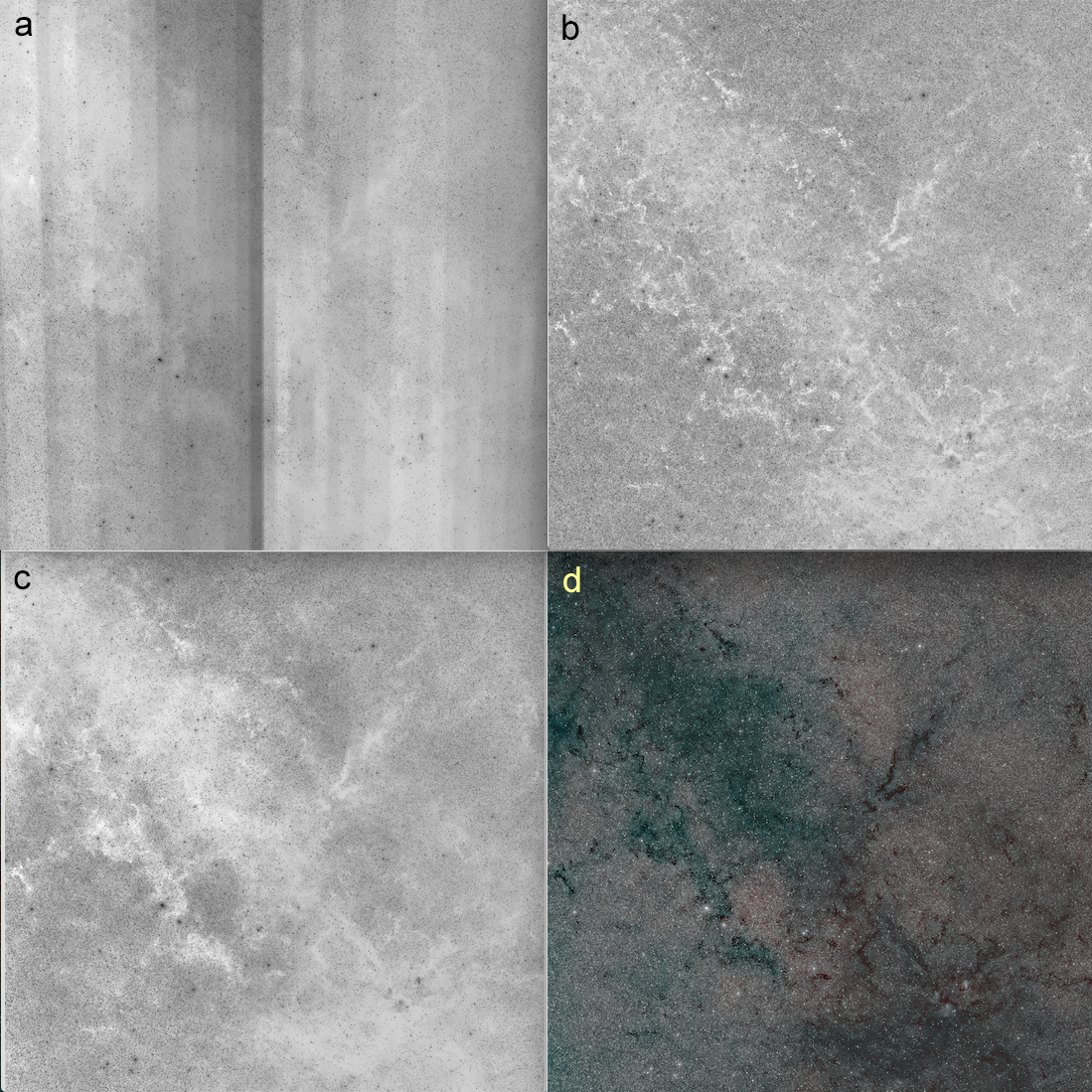}
    \caption{Mosaics of a 5$^o$ x 5$^o$ area in the\ 2MASS  J-band centered  at $\it l$=355$^o$ and $\it b$=0$^o$ and created with Montage to show the effect of background rectification methods. The triangle of stars with some reflection nebulosity
toward the lower right is NGC 6357. (a) No rectification; (b) Flat local background removed; (c) Modeling of the background with Montage; (d) A color map superposing the local background removal, shown as blue/green, and the modeling, shown as red.}
    \label{back}
\end{figure}

Figure~\ref{back} demonstrates the impact of background rectification on the content of an image mosaic constructed from 2MASS images in the J-band. The striped appearance of Figure~\ref{back}(a), where no rectification has been carried out, reveals the background variations across the individual images. In Figure~\ref{back}(b), these background variations have been removed by applying a local flat background from each image. In effect, it acts as a high-pass filter, where the lowest frequency passed is on the scale of an individual input image. This type of filtering is most effective when the image is of a field of sources on a``black sky"; see e.g., \citep{bert02}. Figure~\ref{back}(c) shows the effect of modeling the backgrounds seen in Figure~\ref{back}(a) with the Montage technique. The color map in Figure~\ref{back}(d) superposes the maps in Figures~\ref{back}(b) and 2(c). The linear filter brings out the filamentary structure in the original mosaic, at the expense of showing the large-scale structure of the molecular clouds:  this is made clear in Figure~\ref{back}(d) as the extensive red-colored areas. Thus, the Montage algorithm is most valuable in revealing the large scale structure of an image. All subsequent images in this paper were created with the Montage global relaxation technique.

\section{Using Image Stretches In Montage}
\label{stretchImage}

The pixel values in astronomy images tend to be clumped near the low end of the data range, with a tail at the high end of the data range due to astronomical sources. How can such images be stretched to reveal their faint features without saturating the brighter pixels?  There is no formulaic answer to this question. The optimum stretch is determined by the properties of the image itself and the features an astronomer wishes to emphasize, as well the physiology of the eye and the non-linear response of the monitor. Histogram equalization as commonly used in computer visualization is not useful when applied to astronomical images because it tends to relegate the brightest pixels to a single brightness bin. 

{\tt mViewer} uses an {\it adaptive} histogram equalization algorithm. It assumes the data follow a model where there is a low-level, largely random population containing the majority of the pixels and a long positive tail of more unevenly-spaced bright pixels. There are  two classes of structures at the low level:  random noise and low-level structure made up of faint stars or galaxies. While the origins of these classes are different, they do have similar histograms and can therefore be treated by a common approach. From the image histogram, Montage determines the mean and standard deviation of the low-level distribution, and characterizes data levels in terms ``sigma" values in addition to absolute data values and/or percentiles. Then, rather than base equalization on a uniform target distribution, Montage bases it on the nominal Gaussian distribution, via the error function erf() or via a logarithmically transformed error function. The net result is a histogram where there are a reasonable number of high-level bins showing the brightest areas/pixels, adequate detail at the low end, and adaptive flux-sensitive bins in between. The price paid for this approach is that in order to provide enough bins to show high-level structure, the algorithm may compromise on the number available at the low end. 

This algorithm offers considerable flexibility to astronomers. It optimizes three features at once: the structure of the brightest pixels; the definition of faint structures; and the definition of mid-brightness level structure. Perhaps the most useful feature may be removing the need to carefully choose the high-level cutoff. There is generally no need to use anything but the highest data value as the algorithm maximum. The low-level choice is still manual, and can usually be chosen based on the nature of the background. If the background is all ``noise" (e.g., a field of stars or galaxies) then the low-level pixels can be discarded and a minimum of ``1 $\sigma$" (one standard deviation above the mean background level) generally suffices. If the ``background" is astrophysical (e.g., dense stars in the Galactic plane, maps of clouds of dust and gas, etc.) then ``-2 $\sigma$" is more appropriate. 

\begin{figure} 
    \centering
    \includegraphics[width=6in]{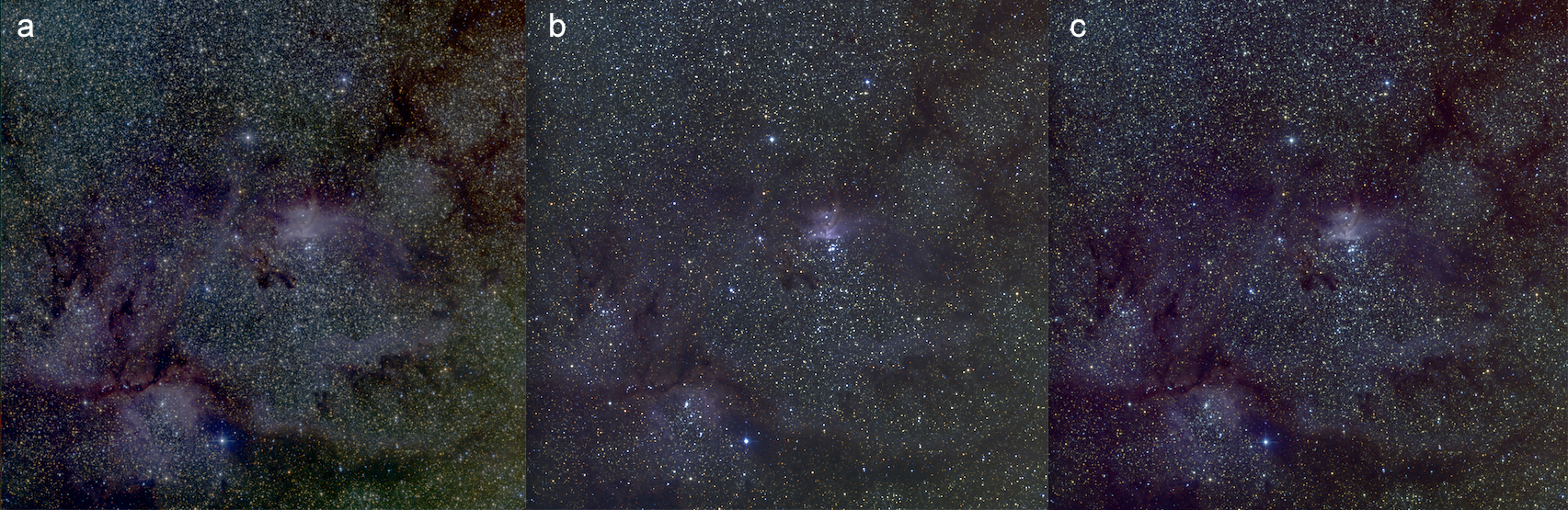}
    \caption{Three 2MASS JHK color composite image mosaics of NCG 6357 shown side-by-side to compare three image presentation algorithms; (a) adaptive histrogram matching used by Montage; (b) a stretch based on the hyperbolic sine function; and (c) application of the gamma correction.}
    \label{stretch}
\end{figure}

It is instructive to compare the adaptive algorithm with two powerful display mechanisms.  \cite{lupt04} have shown the value of using a stretch based on a the hyperbolic sine function, while \cite{bert12} has reasoned that gamma compression and expansion, which reflects the  non-linear luminance of display devices, renders the use of an external stretch function unnecessary. Figure~\ref{stretch} compares the three mechanisms side-by-side. The figure shows how the adaptive algorithm preserves detail across the full range of the image: it shows the structure of the nebulosity as revealed in the gamma compression and the reddening effects in the right corner, as revealed by the hyperbolic sine stretch. 

The hyperbolic sine stretch is very good at revealing detail in many images, particularly those measured by missions such as SDSS, which have galaxies superposed on a dark background.  Because color is sensitive to the gamma correction, this method is well suited to creating color images, especially for rendering on a display device. An image stretching primer at \url{http://bit.ly/2dKlLlP} compares the three methods. It illustrates the above remarks by presenting side-by-side displays of images with different characteristics, including large versions of the images in Figure~\ref{stretch}. Given that the adaptive algorithm preserves well the dynamic range of an image, all the images in the rest of this paper have been created with this technique.

Sometimes, a set of images requires the same stretch. To do this, Montage uses the dedicated utility {\tt mHistogram} to generate a histogram based on a reference image. It uses the same algorithm as {\tt mViewer}, except that it writes the results to a file, which {\tt mViewer} can then use as input for processing a collection of files. \cite{baal16} took advantage of this capability in creating a full-resolution, five-color mosaic of the Herschel Hi-GAL Survey of the Galactic plane \citep{moli10}, for display on the dome of the Fiske Planetarium, Boulder, CO. After creating a mosaic of the Galactic plane at each wavelength, they subdivided the FITS files with {\tt mSubimage} to create more manageable files of size 7,000 x 40,000 pixels, which were converted to PNG files with a common stretch through {\tt mHistogram} and {\tt mViewer}. These were then stitched together with Photoshop to create the final image for display. When complete, the full-dome presentation will cover 360$^o$ by 2$^o$ of the Galactic plane in all five wavelengths. All the data will be processed at full resolution, and can be zoomed on the dome to show details at scales of $\approx$ 10 arcseconds. The images will be presented in monochrome or as color composites. 

\section{Creating Sky Coverage Maps: {\tt mViewer} as a Sky Graphics Engine}
\label{Sky Coverage}

\begin{figure} 
    \centering
    \includegraphics[width=6in]{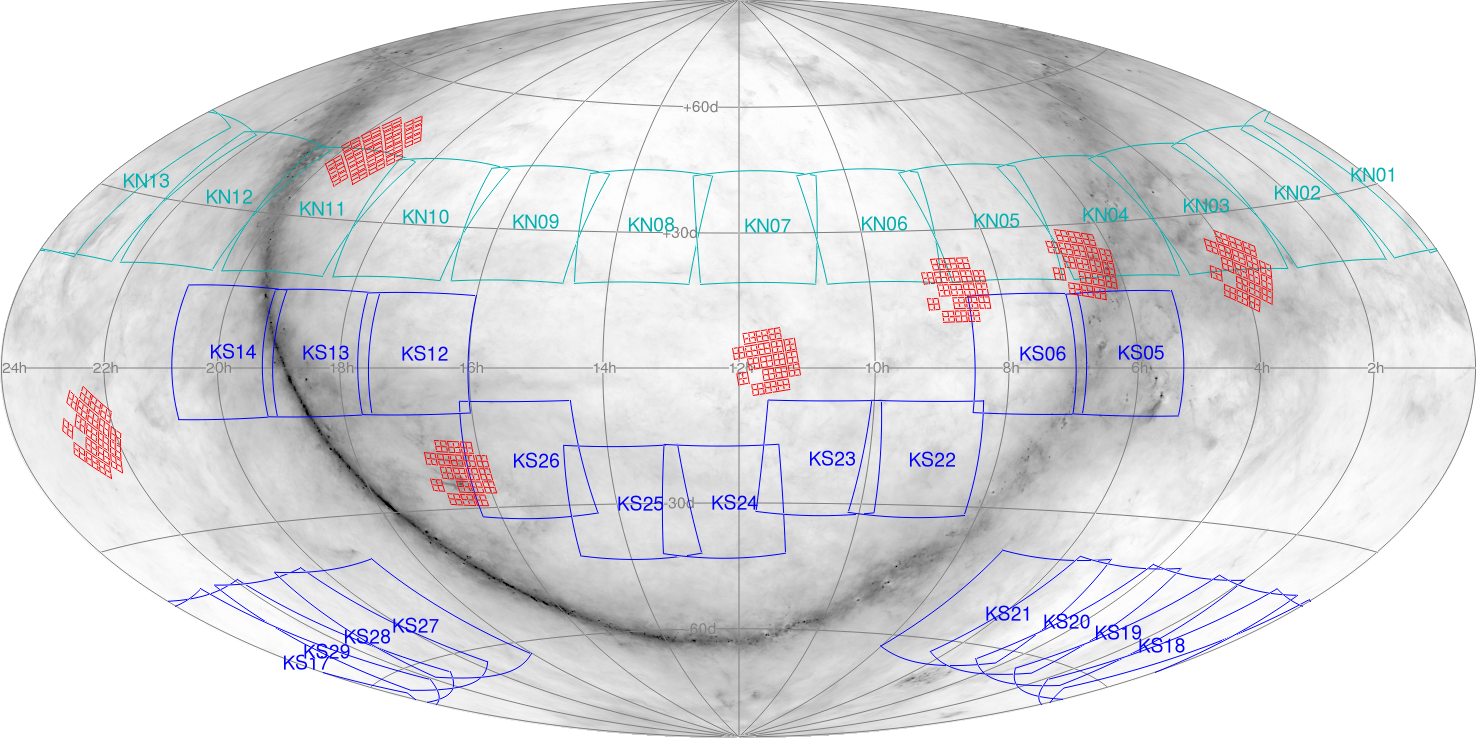}
    \caption{The sky coverage of the KELT-N fields, in turquoise, and the KELT-S field, in blue, compared with those of the Kepler and K2 missions, in red. The fields are shown superposed on the 100$\mu$m map of \cite{schl98}}
    \label{KELT}
\end{figure}

There are instances where the graphical overlays on images are themselves the goal of the visualization. The most common example is to represent image footprints or project coverage footprints on the sky. Figure ~\ref{KELT} shows an example of the coverage on the sky of the Kilodegree Extremely Little Telescope (KELT) survey fields. This map was requested of Montage by the KELT team to visualize the overlap between the KELT fields and the Kepler \citep{boru16} and K2 mission fields \citep{howe14}. KELT surveys the sky for new transiting planets around bright stars, in sets of fields that are 26$^\circ$ x 26$^\circ$ in size. The project operates two observing stations, KELT-North \citep{Pepp07} and KELT-South \citep{Pepp12}. The KELT-North fields are shown in turquoise, and the KELT-South fields in blue, while the Kepler and K2 footprints are shown in red. All these footprints are superposed on a reverse grayscale image of the 100$\mu$m map of \cite{schl98}. To create this image, the Kepler and K2 footprints and the KELT field footprints are written as IPAC (column-delimited) ASCII tables. Supplementary Tutorial 1 shows how to construct the image.

\section{Creating Animations of Data Cubes}
\label{Cubes}

The Montage YouTube channel \url{(https://www.youtube.com/channel/UCFjmHCDrq4YIUly1r082TjA}) shows three animations of mosaics of data cubes of the Galactic Arecibo L-band Feed Array HI (GALFA-HI) survey Data Release 1 (DR1) data set \citep{Peek11}, which covered 13,000$^\circ$ on the sky at 4 arcmin resolution. The mosaics have been created with Version 4 of Montage, and are structured with Right Ascension and Declination in the x- and y-dimensions, and the HI velocity in the z-dimension; altogether there are 2048 velocity planes represented in the z-dimension. The creation of the animations is straightforward. {\tt mViewer} creates a PNG representation of each velocity plane, and the collection of images are input into a video or animation editor. The GALFA animations have been created with the ImageMagick$^{TM}$ suite, but many tools are adequate for this purpose. 

One animation, at \url{https://youtu.be/p2t6Oyw42cg}, shows a full-resolution mosaic of all 2048 frequency planes of 30 GALFA-HI images centered on 0h Right Ascension. A second, at \url{https://youtu.be/Ygu8xLZoK8I}, shows a full-resolution mosaic of the central 256 frequency planes of 30 GALFA-HI images, centered on 0h Right Ascension, with the RGB color derived by combining three adjacent frequency planes. Both mosaics were computed on the Amazon Elastic Cloud 2 (EC2) of Amazon Web Services (AWS) \citep{berr16}, and required five processing hours on a virtual cluster of five machine instances. These processing times would become prohibitive on a desktop, and a simple solution is to create the animations with images that have been downsampled with {\tt mShrink}. The third video, at \url{https://youtu.be/59z_whh0UJI}, was created this way. It represents an average of the central 10 velocity planes of a mosaic of five GALFA data cubes. Supplementary Tutorial 2 describes the creation of this product.

\section{Visualizing Maps in HEALPix Format}
\label{HE}

The HEALPix sky-tessellation scheme is designed to optimize harmonic analysis of wide areas of the sky \citep{gors05}, and has become the standard for recording data acquired by surveys of diffuse background radiation. All HEALPix pixels at a given resolution have the same area and the pixel centers are arranged in latitude bands. Levels of increasing resolution are derived by recursive splitting of these pixels into four equal portions. The cell numbers computed by this scheme are written in a FITS table, rather than as FITS images, and in this form are not suited for visualization. \cite{calro07} have shown that HEALPix FITS tables can be mapped to a hybrid spherical projection class that combines a cylindrical equal area projection at low latitudes with a Collignon projection nearer the poles. The HEALPix pixels become perfect diamonds in this projection, and rotating the image space by 45$^o$ maps the data into a standard pixel array (with the penalty that half the space is empty). Figures 1 and 2 of \cite{calro07} show the projection graphically. The WCSLIB package (\url{http://www.atnf.csiro.au/people/mcalabre/WCS/wcslib/}), which implements the WCS standard, includes a utility, {\tt HPXcvt}, that converts HEALPix FITS tables to FITS pixel images, with a spherical projection identified by ``HPX." This transformation involves no resampling of the data because the image pixels have a one-to-one correspondence with the HEALPix cells in the FITS table. With the HEALPix data now written in a spherical projection in a FITS file, Montage simply treats it as another spherical projection. The WCSTools library has been extended in Montage to support the HPX projection (although users who prefer to use the SAO library will lose HEALPix support). 

Visualization of HEALPix data then becomes straighftorward. The Montage reprojection routines transform the FITS images to the projection desired for visualization, and {\tt mViewer} creates a PNG version of the reprojected image. Two reprojection routines are applicable here. {\tt mProject} redistributes flux from the input to the output pixels on the sky and is guaranteed to conserve flux. A new module in Version 5, {\tt mProjectQL}, uses the Lanczos image interpolation scheme \citep{bubu10}, also used by the SWarp mosaic engine \citep{bert02}, to provide higher performance at the expense of conservation of flux; we recommend {\tt mProjectQL} primarily for creating images for quick-look visualization rather than for science analysis. 

\begin{figure} 
    \centering
    \includegraphics[width=6in]{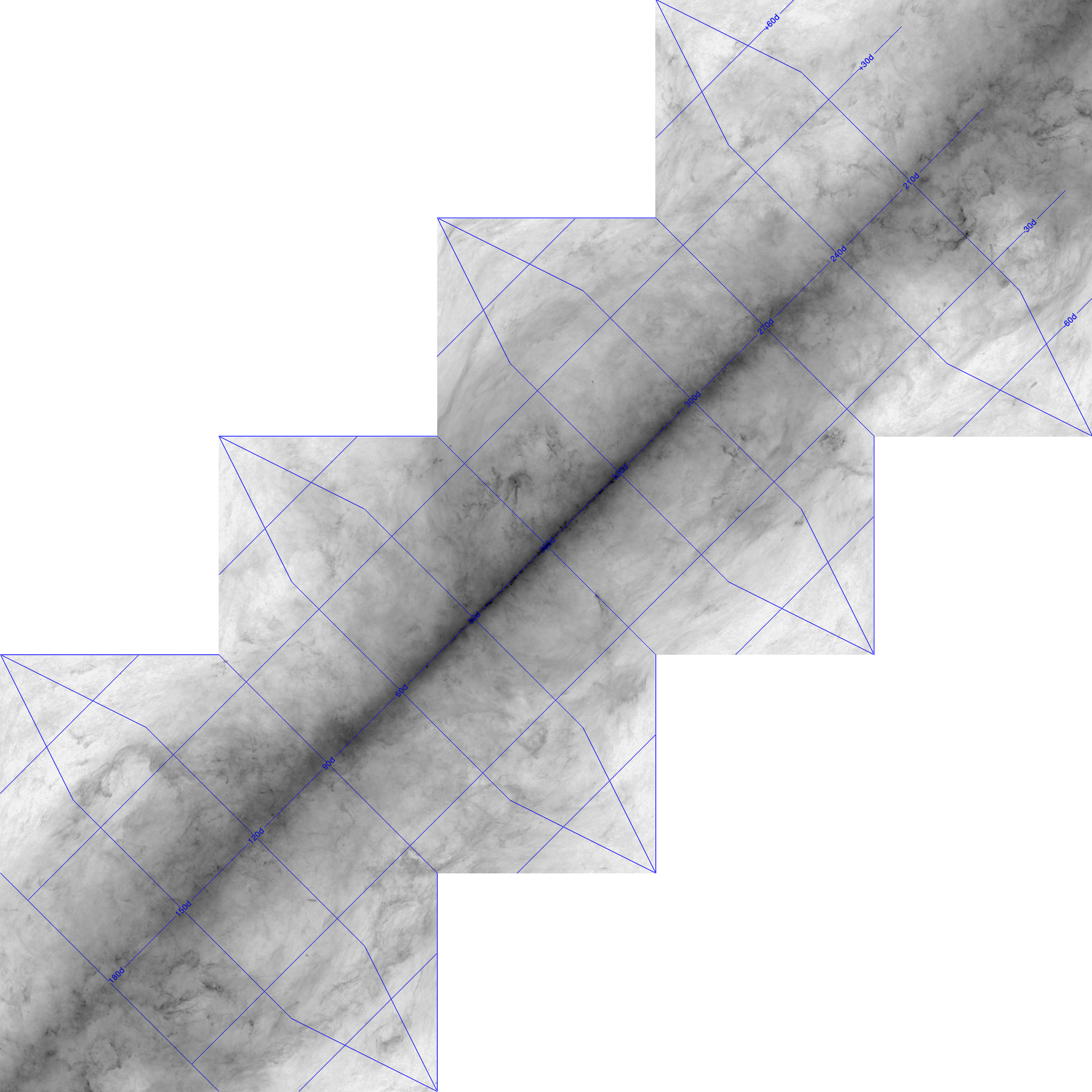}
    \caption{The Planck 857 GHz skymap transformed to the HEALPix projection, as defined by \cite{calro07}, with an Equatorial J2000 grid superposed in blue.}
    \label{PlanckHPX}
\end{figure}

\begin{figure}
    \centering
    \includegraphics[width=6in]{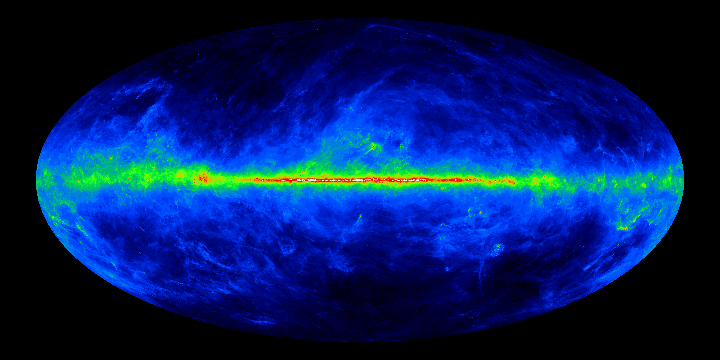}
    \caption{The Planck 857 GHz skymap in Figure ~\ref{PlanckHPX} reprojected to the Aitoff projection by Montage.}
    \label{PlanckAIT}
\end{figure}

\begin{figure}
    \centering
    \includegraphics[width=6in]{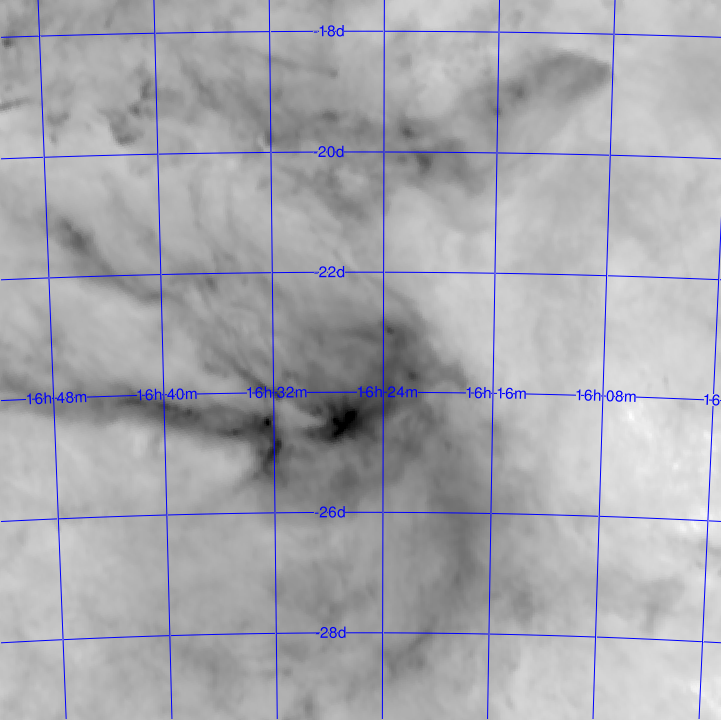}
    \caption{A grayscale image of Rho Ophiuchus region sampled at 1 arcmin, shown in gnomnic projection, with an Equatorial J2000 grid superposed in blue. The image was created with Montage}
    \label{PlanckRhoOph}
\end{figure}

 Figure~\ref{PlanckHPX} shows the Planck All-Sky Map at 857 GHz map in HPX format, downloaded from the NASA/IPAC Infrared Science Archive (\url{http://irsa.ipac.caltech.edu}) and Figure~\ref{PlanckAIT} shows the same map after conversion to an Aitoff projection, which is suitable for displaying all-sky maps. Figure ~\ref{PlanckRhoOph} shows an example of a small region of this map:  Rho Oph in Gnomonic projection image with 1 arcminute pixels in Equatorial coordinates. Supplementary Tutorial 3 shows how to create these images.

\section{Displaying Images in the World Wide Telescope}
\label{Toast}

The WWT is a visualization tool for astronomical data, developed by Microsoft Corporation \citep{good12}. The American Astronomical Society (AAS) assumed responsibility for its management in January 2016, and at the same time released the code with an open-source license. The WWT Windows and web client interfaces allow users to pan and zoom across the sky, and discover and visualize image surveys and pointed observations. 

Version 5 of Montage provides a mechanism for processing astronomy images so  they comply with WWT's special requirements for consuming and displaying images. Images must comply with the TOAST  sky-partitioning scheme (\url{http://www.worldwidetelescope.org/docs/WorldWideTelescopeProjectionReference.html}). Each TOAST pixel is itself a pair of triangles, as defined originally by the Hierarchical Triangular Mesh (HTM) indexing scheme developed by \cite{szal05}. The WWT also imposes requirements on the organization of files for consumption. The data must be JPEG or PNG files 256 x 256 pixels in size. The highest level of these files covers the entire sky. The next level is a set of four images covering longitude quadrants (N-S pairs of HTM octants), and so on to as fine a resolution as is required to display the data. 

As is the case with HEALPix in Section~\ref{HE}, Montage has taken the approach of treating TOAST as another spherical projection, so its reprojection routines can process the image data as they would any spherical projection. Because FITS files containing TOAST projections cannot be consumed by WWT directly, Montage provides a set of dedicated utilities to create properly organized PNG files. Thus, users can create visualizations of the images within WWT without knowledge of the WWT's special requirements. The next two subsections describe in more detail how to prepare the Planck HEALPix maps data at 857 GHz for consumption by the WWT, and Supplementary Tutorial 4 (see Table~\ref{sst}) takes readers through this process step-by-step.

\begin{figure}
    \centering
    \includegraphics[width=6in]{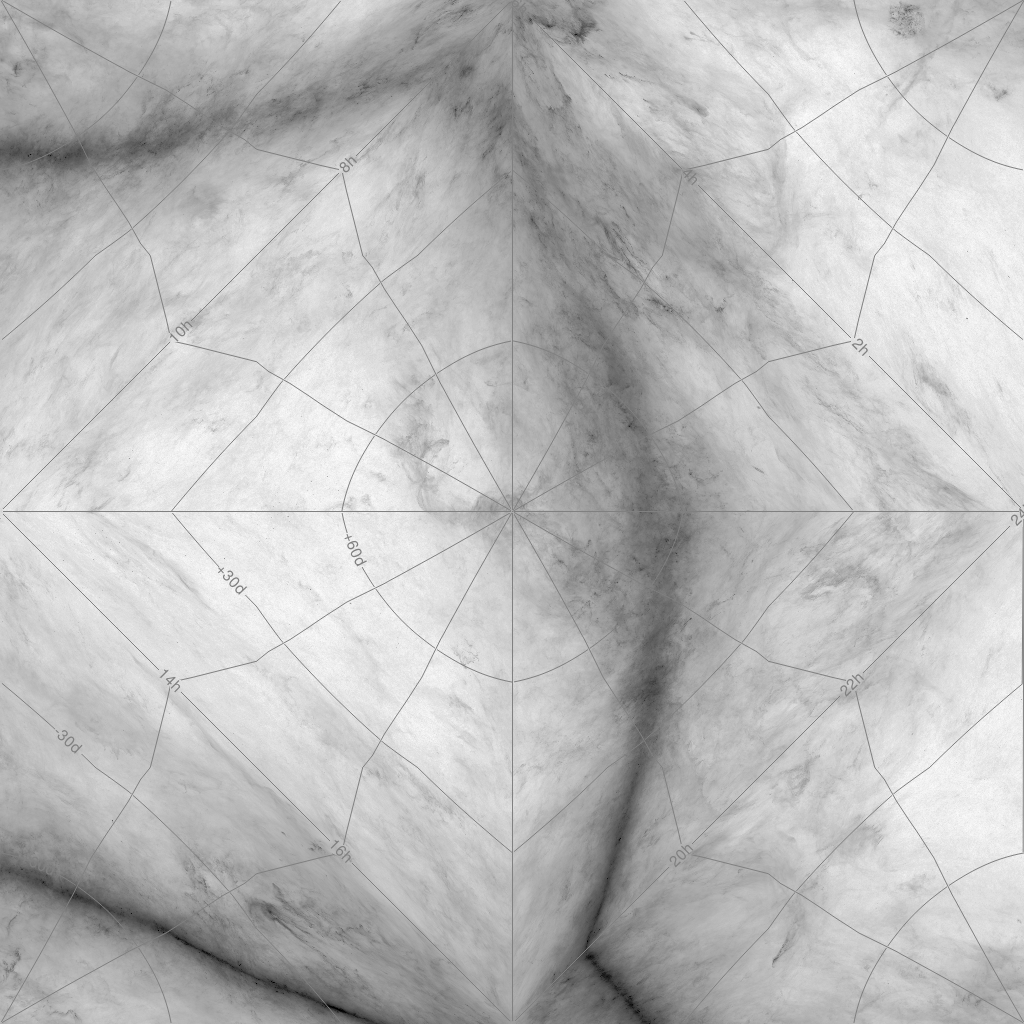}
    \caption{A grayscale image of the Planck 857 GHz all-sky map shown in the TOAST projection, with an Equatorial J2000 grid superposed. See text for a detailed explanation of the TOAST projection as used by Montage.}
    \label{TOAPla}
\end{figure}

\subsection{Creating FITS Images With TOAST As A WCS Projection}

When viewed as a spherical projection, HTM and TOAST differ from standard spherical projections such as gnomonic and simple cylindrical, where the transformation between pixel sky coordinates is formulaic. Determining which HTM cell or TOAST pixel corresponds to a location on the sky requires starting with the base level HTM octant triangles and ``drilling down," finding arc midpoints, connecting them with great circle segments, and determining which subcell in a location is placed.   The TOAST calculations may appear computationally intensive since they involve drilling down from the full sphere to HTM vertices for every pixel corner. This apparent cost is, however,  deceptive bacause the HTM calculations involves a at most a few tens of dot- and cross-products with no trigonometric functions.  In contrast to this, many datasets involve computation of tens of polynomial distortion coefficients in addition to such trigonometric and inverse trigonometric calculations as are needed for the
projection itself.  In practice, the TOAST reprojections turn out to be similar in total compute time as those for many formulaic projections.

Another area where HTM and TOAST differ from standard projections is that because they start with the whole sky and always subdivide the same way, only a discrete set of pixel scales are possible, and this impacts how the WCS parameters are managed and applied. Parameters from WCS like CDELT CD, which ordinarily control the scale of the image, are only informational for TOAST. With HTM, the level parameter controls the image scale and this is captured as keyword PV2\_1. Table~\ref{TOASTHeader} in Appendix~\ref{head} shows the sample FITS header used in the example in Tutorial 4. Montage uses a custom modification of the WCSTools package to support TOAST: users employing the WCS library directly from SAO will lose the TOAST functionality. Images processed in the TOAST  projection,``TOA," are not well suited for direct visualization. Figure~\ref{TOAPla} is a TOAST representation of the 857 GHz Planck sky map described in the last section. The image is mirror-imaged relative to a normal all-sky projection and there are discontinuities in the slopes of curves, best seen in the Right Ascension and Declination lines. 

\subsection{Generating PNG Images from FITS Images In The TOAST Projection}
\begin{figure}
    \centering
    \includegraphics[width=6in]{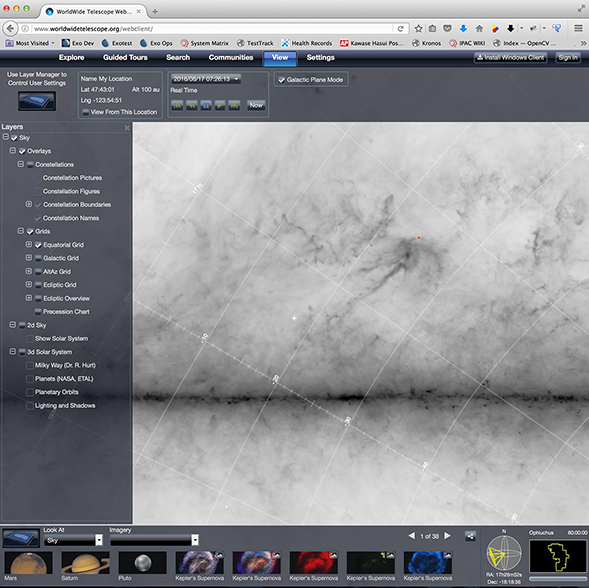}
    \caption{A sample screen shot of the WWT web interface, presenting a section of the Planck 857 GHz map.} 
    \label{WWT}
\end{figure}

The Planck image in Figure~\ref{TOAPla} cannot be consumed by WWT, even when represented as a PNG image,  because it is not organized in the WWT tiling scheme: it shows the whole sky at HTM level two (1024 x 1024 pixels) and contains 4x4 TOAST tiles. Montage therefore contains dedicated utilities for converting a TOAST FITS file to a PNG file, organized and named as WWT requires. WWT expects to find a set of PNG images that are 256 x 256 pixels in size and processes whatever subset it requires for the region and zoom-level it is presenting. It starts with the single all-sky image for level 0, four for level 1, then 16, 256 and 1024 for levels 3, 4 and 5. The Planck example goes as far as level 5, which corresponds to the intrinsic resolution of the Planck original data. So to support consumption by the WWT, Montage must produce a total of 1365 images, each of size 256 x 256 pixels. WWT supports several naming conventions for these files. Montage generates them in a recursive ``Z-order pattern," which gives images names such as  ``Planck.png" (for level zero), then ``Planck2.png," ``Planck23.png," ``Planck232.png" and ``Planck2320.png." Montage includes a dedicated set of utilities to create from a set of input images all the required PNG files, organized and named for consumption by the WWT.  We anticipate that most astronomers will use the web version of WWT. In this case, there are two other steps needed. The PNG files must be copied to a URL-accessible location, and an XML file describing the ``image collection" must also be made web-accessible. Appendix~\ref{samplexml} provides a sample XML file, which can be edited by users. Figure~\ref{WWT} shows the Planck map processed to meet the WWT's requirements and presented in the WWT web interface.

\section{Conclusion}

This paper shows how the Montage image mosaic engine is applicabie to the field of visualization. The scalable toolkit design has allowed integration into visualization systems, usually to reproject and resample images at multiple wavelengths or for presentation in a client or browser. A visualization tool {\tt mViewer} supports automated creation of JPEG and PNG representations of FITS images and integration with Python. It enables the creation of images with overlays with a single call, and includes an adaptive image stretch algorithm that preserves the dynamic range of the image. Visualizations contained in this paper with {\tt mViewer} can be recreated and extended in five online tutorials. A background-matching algorithm that models the variations in sky brightness across a mosaic rectifies the background to a common level and enables the faint structure in an image to be more readily seen. Utilities for creating cutouts and downsampled versions of images are useful for visualizing large images. Version 5 of Montage offers support for visualizing data written in HEALPix sky-tessellation scheme, and functionality for processing and organizing images to comply with the TOAST sky-tessellation scheme, as required for consumption by the WWT. 

\acknowledgements
Montage is funded by the National Science Foundation under Grant Number ACI-1440620, and was previously funded by the National Aeronautics and Space Administration's Earth Science Technology Office, Computation Technologies Project, under Cooperative Agreement Number NCC5-626 between NASA and the California Institute of Technology. The Galactic ALFA HI (GALFA HI) survey data set was obtained with the Arecibo L-band Feed Array (ALFA) on the Arecibo 305m telescope. Arecibo Observatory is part of the National Astronomy and Ionosphere Center, which is operated by Cornell University under Cooperative Agreement with the U.S. National Science Foundation. The GALFA HI surveys are funded by the NSF through grants to Columbia University, the University of Wisconsin, and the University of California. The authors thank the AWS SKA AstroCompute Program for the award of educational credits to support the processing of GALFA data. This research has made use of the NASA/ IPAC Infrared Science Archive, which is operated by the Jet Propulsion Laboratory, California Institute of Technology, under contract with the National Aeronautics and Space Administration. This publication makes use of data products from the Two Micron All Sky Survey, which is a joint project of the University of Massachusetts and the Infrared Processing and Analysis Center/California Institute of Technology, funded by the National Aeronautics and Space Administration and the National Science Foundation. Funding for the creation and distribution of the SDSS Archive has been provided by the Alfred P. Sloan Foundation, the Participating Institutions, the National Aeronautics and Space Administration, the National Science Foundation, the U.S. Department of Energy, the Japanese Monbukagakusho, and the Max Planck Society. The SDSS web site is \url{http://www.sdss.org/}. The Participating Institutions are The University of Chicago, Fermilab, the Institute for Advanced Study, the Japan Participation Group, The Johns Hopkins University, the Max Planck Institute for Astronomy (MPIA), the Max Planck Institute for Astrophysics (MPA), New Mexico State University, Princeton University, the United States Naval Observatory, and the University of Washington. We wish to thank Dr. J. Bally and Mr. J. E. Allured for permission to quote results prior to publication. We thank Dr. B. Rusholme for his collaboration in processing the GALFA data, Dr. J. Pepper for permitting us to use the unpublished Figure ~\ref{KELT}, and Ms. Marcy Harbut for editorial assitance with the manuscript.

\appendix
\section{Sample FITS File When Treating TOAST As A Spherical Projection}
\label{head}

This Appendix presents a sample FITS file contains the header information required by the TOAST projection for subsequent presentation in the WWT.  All the HTM calculations that create values recorded in  this header are computed to a level that is equivalent to a spatial scale of a fraction of a milliarcseconds, adequate for visualization.

\begin{deluxetable}{lll}
  \tablecolumns{3}
  \tablewidth{0pc}
  \tablecaption{\label{TOASTHeader} Sample FITS Header When Treating TOAST As A Spherical Projection}
  \tablehead{
    \colhead{Parameter}  &
    \colhead{Definition}  &
    \colhead{Sample Value}   \\
  }
\startdata
NAXIS     &     Number of axes &        2   \\
NAXIS1   &       Size of axis 1    & 256      \\ 
NAXIS2  &     Size of axis 2 &    256         \\
CTYPE1   &   Name of the coordinate axis 1 &   'RA---TOA'  \\
CTYPE2  &   Name of the coordinate axis 2 &	'DEC--TOA'	 \\
CRPIX1    &    Coordinate system reference pixel along axis 1 &     -3072.50  \\
CRPIX2    &    Coordinate system reference pixel along axis 2  &   -1536.50   \\
PV2\_1     &   Parameter describing image projection   &     5  \\
XTILE       &  Tile coordinates  &      12    \\
YTILE       &   Tile coordinates  &       6  \\
CDELT1    &   Coordinate increment along axis   1 &     1.00   \\ 
CDELT2  &      Coordinate increment along axis 2   &  1.00    \\ 
CRVAL1   &      Coordinate system value at reference pixel  &    0.  \\
CRVAL2  &         Coordinate system value at reference pixel &    0.   \\  
PC1\_1    &      PC matrix element  &      1.00   \\  
PC1\_2   &        PC matrix element  &          0.00  	\\
PC2\_1   &       PC matrix element  &           0.00     \\
PC2\_2   &     PC matrix element  &       1.00    \\
\enddata
\label{TOA}
\end{deluxetable}

Notes: 

 \begin{itemize}
 
 \item Keywords of the type PV\_m were introduced into the FITS standard to take account of non-linear parameter values for those projections that required them \citep{grca02}, and usage is custom to the projection in use. In the case of TOAS ,  PV2\_1 is used here to describe the HTM level; that and the requirements of the TOAST file organization scheme drive the values of the keywords in the header. 
 \item The TOAST tiles for consumption by the WWT are always 256 x 256 pixels in size and are arranged in a regular XY array. We have included the "tile coordinates" in the parameters XTILE and YTILE, though these are not used in the computation. They are for informational use only.
 \item The parameters CDELT, CRVAL, and the PC matrix are all fixed boilerplate values, but the Montage instance of WCSTools requires that they are present.
 \item  The CRPIX values represent the pixel offset from the first pixel in the file and the edge of the "untiled" image for this HTM level (e.g., -256 * XTILE - 0.5).

\end{itemize}

\section{Sample XML Template for Describing Image Collections For Consumption by the WWT}
\label{samplexml}

The WWT web interface requires an XML file describing the set of files for consumption them. This sample XML file can be edited by users to descibe their own collections.  

\begin{verbatim}

<?xml version="1.0" encoding="UTF-8"?>

<Folder Name="Montage Tests"
        Group="Explorer"
        Searchable="True"
        Type="Sky"
        Thumbnail="http://montage.ipac.caltech.edu/workspace/Planck/icon/color_AIT_small.png">


  <ImageSet Generic="False"
            DataSetType="Sky"
            BandPass="microwave"
            Name="Planck 857 GHz"
            Url="http://montage.ipac.caltech.edu/workspace/Planck/857/Planck{Q}.png"
            BaseTileLevel="0"
            TileLevels="5"
            BaseDegreesPerTile="180"
            FileType=".png"
            BottomsUp="False"
            Projection="Toast"
            QuadTreeMap="0123"
            CenterX="0"
            CenterY="0"
            OffsetX="0"
            OffsetY="0"
            Rotation="0"
            Sparse="False"
            ElevationModel="False"
            StockSet="False">

    <ThumbnailUrl>
       http://montage.ipac.caltech.edu/workspace/Planck/icon/Planck857_AIT_small.png
    </ThumbnailUrl>

    <Credits>
       Montage reverse grayscale example made from Planck HEALPIX 857 GHz data.
    </Credits>

    <CreditsUrl>
       http://montage.ipac.caltech.edu/
    </CreditsUrl>

  </ImageSet>

</Folder>

\end{verbatim}

\allauthors

\end{document}